\documentclass[aps,pra,twocolumn,groupedaddress,showpacs,10pt]{revtex4-1}
\usepackage{hyperref}

\begin{document}


\title{Evidence procedure for efficient quantum state tomography}


\author{Jochen Rau}
\email[]{jochen.rau@q-info.org}
\homepage[]{www.q-info.org}
\affiliation{Institut f\"ur Theoretische Physik,
Johann Wolfgang Goethe-Universit\"at,
Max-von-Laue-Str. 1, 
60438 Frankfurt am Main,
Germany}


\date{\today}

\begin{abstract}
I show that in tomographic experiments measurement of a small set of observables suffices to confirm or incrementally amend prior expectations with a high degree of confidence.
To this end I adapt the evidence procedure, an estimation technique used in classical image reconstruction, to use in quantum state tomography.
\end{abstract}

\pacs{03.65.Wj, 02.50.Cw, 02.50.Tt, 05.30.Ch}

\maketitle

\section{\label{intro}Introduction}

The reconstruction of a quantum state from experimental data is generally imprecise for a variety of reasons.
Measurement devices work with limited accuracy;
the sample size is finite;
and the set of observables used might not be informationally complete.
Rather than singling out a quantum state uniquely, such imperfect data are generally compatible with an infinite set of different states. 
Identifying the most plausible member of this set can be a formidable challenge that requires advanced statistical estimation techniques \cite{dariano:tomographic}.
For this task there exist numerous theoretical schemes,
some widely used examples being pattern functions \cite{dariano:pattern,leonhardt:pattern}, maximum likelihood \cite{hradil:estimation,james:qubits,hradil:lnp} or -- in case of informational incompleteness -- the maximum entropy method \cite{buzek:spin,buzek:aop,buzek:reconstruction,buzek:lnp}.

Yet while, e.g., maximum likelihood estimation has proven successful in many practical applications, it reaches a theoretical limit and may even lead to inconsistencies when applied to special situations:
For instance, it may return zero eigenvalues for a density matrix when, in fact, this is just an artefact caused by a small sample size \cite{blume-kohout:optimal}.
This suggests that prior knowledge beyond the ``naked'' data may be important, and has motivated Bayesian modifications to the conventional scheme \cite{blume-kohout:hedged}.
Another shortcoming of many algorithms is that they do not quantify error bars, a piece of information that would be crucial to assess the reliability of an estimate;
here, too, important theoretical developments are currently underway \cite{rehacek:errorbars,audenaert:kalman}.

In addition to their conceptual limitations most reconstruction schemes run into practical difficulties as soon as quantum systems become more complex, as the requirements for experimental and computational resources grow exponentially with the number of constituents of a composite system.
For example, H{\"a}ffner et al. \cite{haeffner:scalable} report that, to reconstruct an entangled state of eight calcium ions, they needed to perform 656,100 experiments.
This has triggered an intense search for alternative quantum tomography protocols that can do with fewer resources \cite{aaronson:learnability}.

Recently a number of proposals have been made for efficient  protocols whose resource requirements grow only polynomially with system size.
All these proposals are limited in scope:
Either they do without complete state reconstruction and instead focus  on ascertaining  some specific property of the state 
such as  the presence or absence of entanglement \cite{audenaert+plenio,wunderlich:incomplete};
or else they presuppose that the unknown state lie in some lower-dimensional reconstruction subspace \cite{rehacek:compatible} or
belong to some privileged subclass of states
such as matrix product states,
with a number of parameters which is only polynomial in system size \cite{cramer+plenio,flammia:polynomial,landon+poulin}. 
Protocols of the latter type  deliver one or both of the following:
(i)
the information whether or not an unknown state does indeed belong (within some given error bound) to the privileged subclass;
and
(ii)
if so, an estimate for the associated state parameters.
While this falls short of  tomography for {arbitrary} states it is nevertheless of great use for many real-world situations
as, e.g., matrix product states capture the low energy physics of a wide range of one-dimensional systems \cite{landon+poulin}.

The recent proposals illustrate that efficiency gains can  be achieved if one makes clever use of \textit{prior knowledge} about the system ---
such as in the above example, the well-founded expectation that the system is described by a matrix product state.
Prior knowledge thus plays a pivotal, and often underappreciated, role in quantum state tomography.
Its use in  a tomographic scheme not only  ensures internal consistency and in particular avoids the zero-eigenvalue problem encountered in conventional maximum likelihood;
but it is also  key to realizing  sorely needed efficiency gains.
In most experimental settings such prior knowledge is in fact available, in the form of well-founded (implicit or explicit) expectations as to the output state.
These expectations might be
based on a theoretical model of the underlying physics,  past  results of the same or similar experiments, or  some combination thereof;
and they may specify an anticipated output state uniquely, or merely its parametric form.
 
The situation where prior expectations   merely favor a certain subclass of states, without any  bias as to the parameter values within this subclass,
can in general be reduced to the case where one  anticipates some unique output state.
For as long as there is a polynomial-time algorithm for finding within the subclass the member which best matches some given experimental data
-- as is the case for matrix product states --
one can always pre-process the data to obtain the best match within the subclass, and subsequently consider this best match to be the unique anticipated  state.
In many realistic cases, therefore, one may assume that 
prior knowledge comes in the form of some unique anticipated  state.

While  there is thus an expectation as to the output state, it is obviously not guaranteed that precisely this anticipated  state
will  be produced in the   experiment ---
after all, if this were certain a priori, what would be the point of the experiment?
The anticipated state  comes only with a finite (and often not precisely quantified) degree of confidence:
It is well possible that due to unwarranted approximations in the theoretical model
or experimental inaccuracies the actual output state  will deviate from the anticipated state.
This deviation is not expected to be very large, for else there would be a serious problem with the theory or experimental setup;
but a quantitative bound  is difficult to establish a priori. 
Strictly speaking, therefore, the prior expectation is that the actual state lies within some unspecified, yet not too large
neighborhood of the anticipated state.
Often the precise magnitude of this deviation  constitutes one of the novel insights provided by the experiment.

In this generic setting  quantum state tomography reduces to  the following tasks:
(i)
verify that the unknown state of the system is indeed in the proximity of (but not necessarily identical with) the anticipated state;
(ii)
whenever it deviates from the anticipated state, prescribe how this initial estimate must be amended;
and 
(iii) quantify the  error bars associated with the updated estimate.
The recent proposals mentioned above accomplish these tasks  in part:
They do supply efficient protocols, yet only for the special case where
the anticipated state is of matrix product form,
and only for the first of the three tasks.
Against this backdrop, the purpose of the present paper is 
to propose a   universal protocol with a significantly broader scope:
a protocol which takes into account prior knowledge in a generic way, allowing arbitrary anticipated states,  
and which  tackles efficiently all three  tasks at once.

In order to achieve this objective I will borrow powerful techniques 
from the classical field of image reconstruction \cite{pryce+bruce,sivia:book}.
There, too, a large amount of noisy (pixel) data needs to be processed with sophisticated algorithms to yield the most plausible tomographic image.
Like in quantum state tomography, one key to improving the efficiency and accuracy of these algorithms has been to use whatever prior knowledge is available, however qualitative or vague.
Mathematically, this is achieved by modelling  prior knowledge with distributions that contain themselves unknown parameters;
these so-called \textit{hyperparameters} are then themselves subjected to an estimation procedure.
In its most general form such a scheme is known as \textit{hierarchical Bayes}.
Practical calculations usually involve well-controlled approximations, the most important being the \textit{evidence procedure} \cite{gull:developments,skilling:displays,skilling:reconstruction,mackay:interpolation,strauss:alpha,hanson:intro,mackay:hyperparameters,wolpert+strauss,mackay:comparison}.
It entails a maximum likelihood approximation applied to the estimation of hyperparameters and is therefore also called \textit{generalized maximum likelihood} or \textit{ML-II}.

Transferring this idea to quantum state tomography, the unknown degree of confidence as to the anticipated state becomes a hyperparameter.
This hyperparameter can only be estimated a posteriori based on the overall compatibility of experimental data with the anticipated state.
The estimation involves a generalized  maximum likelihood approximation which, as it pertains to a one-dimensional distribution only,
is usually of good quality ---
more so than the conventional maximum likelihood approximation which pertains to a higher-dimensional, and hence generally broader, distribution on state space.
The thus estimated hyperparameter establishes whether or not the unknown state of the system is indeed in the proximity of the anticipated state;
and if so, it determines the optimal weight to be attributed to anticipated state and experimental data, respectively,
yielding a posterior estimate that interpolates between the two.
The procedure is optimal in the sense that it maximizes the degree of confidence about the resulting posterior estimate;
and it is efficient in that it does so with high accuracy even when measurements are not informationally complete.

In the present paper I flesh out these ideas in mathematical detail and demonstrate their use in a concrete example.
I shall first adapt the framework of the classical evidence procedure to its novel use in quantum state tomography.
I will establish the criteria for its applicability and show that it is in fact particularly suited for tomography of large quantum systems.
I will further show that the evidence procedure is internally consistent and in particular avoids the zero-eigenvalue problem encountered in conventional maximum likelihood; 
that it  greatly speeds up tomography (certification or incremental amendment of an arbitrary anticipated state) whenever the experimental data are sufficiently close to  prior expectations overall;
and that it provides error bars in a straightforward fashion.
I will then illustrate its use in a simple four-qubit system.

For simplicity I shall disregard inaccuracies introduced by the experimental apparatus and instead focus on the two other sources of imprecision, finite sample size and lack of informational completeness. 
Incorporating experimental errors into the formalism is a straightforward exercise that will be dealt with in future work.

In Section \ref{procedure} I introduce the evidence procedure, already in a mathematical form adapted to quantum state tomography.
In Section \ref{application} I illustrate its use with a toy example, a four-qubit system with parameters tuned (for simplicity) such that all calculations can be done analytically.
In Section \ref{summary} I conclude with a brief discussion.

\section{\label{procedure}Evidence procedure}

Suppose one wants to infer the state $\rho$ of a $d$-dimensional quantum system
from a set of $r$ sample means $\{g_a\}$ measured on a sample of size $N$.
Since $N$ is always finite, and the set of observables $\{G_a\}$ need not be informationally complete,
these data do not specify $\rho$ uniquely but only yield some probability distribution $\mbox{prob}(\rho|N,\{g_a\})$.
If this distribution is sharply peaked at some $\rho_e$ then this $\rho_e$ constitutes a plausible estimate for $\rho$;
and the width of the peak is a measure for the quality of the estimate.
Due to Bayes' rule the post-measurement distribution is the product of likelihood and prior,
\begin{equation}
	\mbox{prob}(\rho|N,\{g_a\}) \propto \mbox{prob}(\{g_a\}|N,\rho)\cdot \mbox{prob}(\rho)
	.
\label{posteriorstate}
\end{equation}
The conventional method of maximum likelihood disregards the prior and focuses on the peak of the likelihood function.
In contrast, the evidence procedure (as do other Bayesian methods) seeks to encode in the prior whatever ancillary information is available, however qualitative and vague, and use this to improve the accuracy and efficiency of the estimate.

Often prior knowledge comes in the form of a bias towards some anticipated state $\sigma$ derived from theoretical considerations or past experience.
If this is the only information available then one expects the prior to be peaked at, and symmetric around, $\sigma$;
yet the degree of confidence as to this bias, and hence the width of the distribution, are generally not known. 
Based on very general considerations of classical statistical inference \cite{skilling:axioms} which readily carry over to the quantum case,
the prior must be a monotonically decreasing function $f$ of the distance of $\rho$ from $\sigma$, measured in terms of the relative entropy:
\begin{equation}
	\mbox{prob}(\rho) = f[S(\rho\|\sigma)]
	,
\end{equation}
where $S(\rho\|\sigma):=\mbox{tr}(\rho\ln\rho-\rho\ln\sigma)$.
In the evidence procedure the prior is in fact modelled as an integral
\begin{equation}
	\mbox{prob}(\rho) = \int_0^\infty d\alpha\, \mbox{prob}(\rho|\alpha)\cdot \mbox{prob}(\alpha)  
\end{equation}
where
\begin{equation}
	\mbox{prob}(\rho|\alpha) = Z(\alpha)^{-1} \exp[-\alpha S(\rho\|\sigma)]
\label{proprhoalpha}
\end{equation}
with (in the Gaussian approximation) 
$Z(\alpha)\approx (2\pi/\alpha)^{n/2}$ and $n=d^2-1$.
The prior thus features an unknown \textit{hyperparameter} $\alpha$ whose value is to be estimated a posteriori.

Given the experimental data the distribution for $\alpha$ is updated according to Bayes' rule
\begin{equation}
	\mbox{prob}(\alpha|N,\{g_a\}) \propto \mbox{prob}(\{g_a\}|N,\alpha)\cdot \mbox{prob}(\alpha)
	.
\end{equation}
While the prior $\mbox{prob}(\alpha)$ is generally broad, reflecting the vagueness of the initial bias, the likelihood function may be peaked around some $\alpha_0$ at which
\begin{equation}
	\left.\frac{\partial}{\partial\alpha}\right|_{\alpha_0} \mbox{prob}(\{g_a\}|N,\alpha) =0
	.
\label{maximum}
\end{equation}
Whenever this peak is sufficiently sharp,
\begin{equation}
	-\alpha_0^2 \left.\frac{\partial^2}{\partial \alpha^2}\right|_{\alpha_0} \ln \mbox{prob}(\{g_a\}|N,\alpha) \gg 1
	,
\label{criterion}
\end{equation}
it is justified to make the approximation 
\begin{equation}
	\mbox{prob}(\rho) \to \mbox{prob}(\rho|\alpha_0)
	.
\label{ml2}
\end{equation}
This amounts to a maximum likelihood approximation which, in contrast to the conventional approach, no longer pertains to the estimation of the full quantum state but to that of the hyperparameter $\alpha$ -- whence the method's alternative names \textit{generalized maximum likelihood} or \textit{ML-II}.

Whenever the approximation (\ref{ml2}) is justified 
state estimation can be based on the posterior
\begin{equation}
	\mbox{prob}(\rho|N,\{g_a\},\alpha_0) \propto \mbox{prob}(\{g_a\}|N,\rho)\cdot \mbox{prob}(\rho|\alpha_0)
	.
\label{posteriorstate2}
\end{equation}
Thanks to the quantum Stein lemma \cite{hiai+petz,ogawa+nagaoka,petz:book,brandao+plenio} the likelihood function can for large $N$ be written in the form
\begin{equation}
	\mbox{prob}(\{g_a\}|N,\rho) \propto \exp[-N S(\mu^\rho_g\|\rho)]
\label{likelihood}
\end{equation}
where $\mu^\rho_g$ is the unique state that minimizes the relative entropy $S(\mu\|\rho)$ with respect to $\rho$
while yielding as expectation values of $\{G_a\}$ the observed sample means $\{g_a\}$.
In the Gaussian approximation it is
\begin{equation}
	S(\mu^\rho_g\|\rho) \approx  S(\rho\|\mu^\rho_g) \approx S(\mu^\sigma_{g(\rho)}\|\mu^\sigma_g)
\end{equation}
where $\mu^\sigma_{g(\rho)}$ and $\mu^\sigma_g$ now minimize the relative entropy with respect to the initial bias $\sigma$ rather than $\rho$,
while yielding for $\{G_a\}$ the same expectation values as $\rho$, $\{g_a(\rho)\}$, or the observed sample means $\{g_a\}$, respectively.
The various $\mu$'s all have a generalized canonical form \cite{ruskai:minrent}:
e.g., 
\begin{equation}
	\mu^\sigma_g \propto \exp\left[ (\ln\sigma-\langle\ln\sigma\rangle_\sigma)- \sum_{a=1}^r \lambda^a G_a\right]
\label{maxent}
\end{equation}
with 
Lagrange parameters $\{\lambda^a\}$ adjusted such that the state yields as expectation values of $\{G_a\}$ the observed sample means.
For a typical bias towards total ignorance, $\sigma=I/d$, this state coincides with the estimate based on maximum entropy \cite{buzek:reconstruction}.

The second factor in the posterior (\ref{posteriorstate2}) is the prior (\ref{proprhoalpha}).
Like the likelihood function it features in the exponent a relative entropy, which can be decomposed into
\begin{equation}
	S(\rho\|\sigma)=S(\rho\|\mu^\sigma_{g(\rho)}) + S(\mu^\sigma_{g(\rho)}\|\sigma)
	.
\end{equation}
Summing up the various terms in the exponents of likelihood and prior, and exploiting the quasi-linearity
\begin{equation}
	(1-t)S(\rho\|\mu)+t S(\rho\|\sigma) = S(\rho\|\rho(\mu,\sigma;t)) + C(\mu,\sigma;t)
\end{equation}
where $\rho(\mu,\sigma;t)\propto \exp[(1-t)\ln\mu+t\ln\sigma]$ and $C$ is independent of $\rho$,
the posterior finally acquires the form
\begin{eqnarray}
	&& \mbox{prob}(\rho|N,\{g_a\},\alpha_0) \propto \nonumber \\
	&& \ \exp [ -\alpha_0 S(\rho\|\mu^\sigma_{g(\rho)}) - (\alpha_0+N) S(\mu^\sigma_{g(\rho)}\|\rho_e) ]
	.
\label{posterioralpha0}
\end{eqnarray}
It is peaked around
\begin{equation}
	\rho_e \propto \exp\left[\frac{\alpha_0}{\alpha_0+N}\ln\sigma + \frac{N}{\alpha_0+N}\ln\mu^\sigma_g \right]
	,
\label{evidenceestimate}
\end{equation}
with error bars of the order $O(1/\sqrt{\alpha_0+N})$ as to the degrees of freedom that have been measured,
and $O(1/\sqrt{\alpha_0})$ as to those that have not.
The estimate $\rho_e$ has the same parametric form as the (generalized) Maxent estimate (\ref{maxent}), yet with Lagrange parameters rescaled by a factor $N/(\alpha_0+N)$.
The evidence procedure thus interpolates between the initial bias and the experimental data:
For $N\ll \alpha_0$ quantum state estimation is dominated by the prior, and one is advised to stick to the initial bias. 
For small sample sizes the zero-eigenvalue problem is thus avoided.
At the opposite extreme, $N\gg \alpha_0$, the rescaling factor approaches unity;
hence as one would expect, quantum state estimation is then dominated by the likelihood function (\ref{likelihood}) and leads to the (generalized) Maxent estimate (\ref{maxent}).

The optimal value $\alpha_0$ for the hyperparameter is yet to be determined, and the validity of the (generalized) maximum likelihood approximation yet to be established.
As for the former, one introduces the marginalization
\begin{equation}
	\mbox{prob}(\{g_a\}|N,\alpha) = \int d\rho\ \mbox{prob}(\{g_a\}|N,\rho)\cdot \mbox{prob}(\rho|\alpha)
\end{equation}
and uses
\begin{equation}
	({\partial}/{\partial\alpha}) \mbox{prob}(\rho|\alpha) = [{n}/{2\alpha}-S(\rho\|\sigma)] \cdot\mbox{prob}(\rho|\alpha)
\end{equation}
to derive from the extremum condition (\ref{maximum}) an implicit equation for $\alpha_0$:
\begin{equation}
	\langle \alpha_0 S(\rho\|\sigma)\rangle_{\alpha_0} = n/2
	,
\end{equation}
where $\langle \ldots \rangle_{\alpha_0}$ denotes the expectation value calculated with the posterior (\ref{posterioralpha0}).
In the Gaussian approximation this expectation value can be decomposed into three summands
\begin{eqnarray}
	\langle \alpha_0 S(\rho\|\sigma)\rangle_{\alpha_0} 
	&=& 
	\langle \alpha_0 S(\rho\|\mu^\sigma_{g(\rho)})\rangle_{\alpha_0} 
	+ \langle \alpha_0 S(\mu^\sigma_{g(\rho)}\|\rho_e)\rangle_{\alpha_0} 
	\nonumber \\
	&& 
	+ \alpha_0 S(\rho_e\|\sigma)
\end{eqnarray}
yielding, respectively,
\begin{equation}
	\langle \alpha_0 S(\rho\|\mu^\sigma_{g(\rho)})\rangle_{\alpha_0} =(n-r)/2
	,
\end{equation}
\begin{equation}
	\langle \alpha_0 S(\mu^\sigma_{g(\rho)}\|\rho_e)\rangle_{\alpha_0} = \alpha_0/(\alpha_0+N)\cdot r/2
	,
\end{equation}
\begin{equation}
	\alpha_0 S(\rho_e\|\sigma) = \alpha_0 N^2/(\alpha_0+N)^2\cdot S(\mu^\sigma_g\|\sigma)
	,
\end{equation}
provided the $r$ sample means are independent.
This leads to the explicit formula
\begin{equation}
	\alpha_0=(1-N_{\rm{min}}/N)^{-1}\cdot N_{\rm{min}}
\label{formulaalpha0}
\end{equation}
with 
\begin{equation}
	N_{\rm{min}}:=r/2 S(\mu^\sigma_g\|\sigma)
	.
\end{equation}

In order for the evidence procedure to be applicable
the hyperparameter must be non-negative, and hence the sample size must exceed $N_{\rm{min}}$.
In addition, the inequality (\ref{criterion}) must hold.
By a similar reasoning as above the latter translates into an inequality for the variance of the relative entropy,
\begin{equation}
	\mbox{var}(\alpha_0 S(\rho\|\sigma)) \ll n/2-1
	,
\end{equation}
the variance again being evaluated in the posterior (\ref{posterioralpha0}).
As before, in the Gaussian approximation this variance can be decomposed into three summands
\begin{eqnarray}
	\mbox{var}(\alpha_0 S(\rho\|\sigma)) 
	&=& 
	\mbox{var}(\alpha_0 S(\rho\|\mu^\sigma_{g(\rho)}))
	+ \mbox{var}(\alpha_0 S(\mu^\sigma_{g(\rho)}\|\rho_e))
	\nonumber \\
	&&
	+ 4\alpha_0^2 \langle ([\sigma-\rho_e];[\rho-\rho_e])_{\rho_e}^2 \rangle_{\alpha_0}
\end{eqnarray}
where in the last summand $(\cdot;\cdot)_{\rho_e}$ denotes the scalar product induced by the entropy \cite{balian:physrep,skilling:classic}, evaluated at ${\rho_e}$.
These three terms yield, respectively,
\begin{equation}
	\mbox{var}(\alpha_0 S(\rho\|\mu^\sigma_{g(\rho)})) = (n-r)/2
	,
\end{equation}
\begin{equation}
	\mbox{var}(\alpha_0 S(\mu^\sigma_{g(\rho)}\|\rho_e)) = \alpha_0^2/(\alpha_0+N)^2 \cdot r/2
	,
\end{equation}
\begin{equation}
	4\alpha_0^2 \langle ([\sigma-\rho_e];[\rho-\rho_e])_{\rho_e}^2 \rangle_{\alpha_0} = 2 \alpha_0^2 N^2/(\alpha_0+N)^3 \cdot S(\mu^\sigma_g\|\sigma)
	.
\end{equation}
Inserting for the hyperparameter the formula (\ref{formulaalpha0}) then gives the criterion
\begin{equation}
	r/2 \cdot (1-N_{\rm{min}}/N)^2 \gg 1
	;
\label{rlarge}
\end{equation}
i.e., the evidence procedure presupposes that the set of observables is large, $r \gg 1$.
Thus the evidence procedure is adapted in particular to the tomography of  quantum systems that are large.

Combining the last inequality once more with the formula (\ref{formulaalpha0}) for the optimal hyperparameter shows that the latter must be inside the range
\begin{equation}
	N_{\rm{min}} < \alpha_0 \ll \sqrt{r/2} \cdot N_{\rm{min}}
\end{equation}
and hence approximately of the same size as $N_{\rm{min}}$.
Consequently, fluctuations {in hitherto \textit{unmeasured} degrees of freedom} are effectively estimated to be of the order $O(1/\sqrt{N_{\rm{min}}})$; 
the hitherto unmeasured degrees of freedom are estimated to an accuracy \textit{as if} they had been measured on a sample of size $N_{\rm{min}}$.
This fictitious sample size $N_{\rm{min}}$ becomes large whenever $S(\mu^\sigma_g\|\sigma)$ is small, i.e., whenever the experimental data are in good agreement with prior expectations overall.
In this case the data, albeit informationally incomplete, provide a high degree of confidence not only as to the degrees of freedom actually measured
but also as to those that have {not}.
If one  seeks confirmation or only incremental amendment of prior expectations, therefore, it is not necessary to measure a set of observables that is informationally complete.

\section{\label{application}Example: Four qubits}

In this Section I apply the evidence procedure to incomplete tomography of a four-qubit system.
The example is intentionally simplified in order to make all calculations tractable analytically.

An $M$-qubit system (say, $M$ entangled photons) is described in a Hilbert space of dimension $d=2^M$.
Complete state tomography would require measurement of $n=2^{2M}-1$ different observables,
i.e., demand resources that grow exponentially with the number of qubits.
As $M$ increases, this quickly becomes impossible in real experiments.
Here I suppose that in an actual experiment one can only determine sample means $\{\vec{s}_i\}$ of the single-qubit spin vectors
\begin{equation}
	\vec{S}_i:=1/2\cdot (X_i,Y_i,Z_i)
\end{equation}
where $X_i,Y_i,Z_i$ are the Pauli matrices pertaining to qubit $i$,
and sample means $\{\mathbf{c}_{ij}\}$ of the qubit-qubit correlation matrices with components
\begin{equation}
	C^{ab}_{ij}:= S^a_i S^b_j - s^a_i s^b_j
	\ ,\ 
	a,b=1\ldots 3
	\ ,\ 
	i\neq j
	.
\end{equation}
Together these are $r=3M^2$ independent sample means,
a number growing only quadratically with the number of qubits.
For four qubits it is $r=48$, less than $20\%$ of an informationally complete set ($n=255$),
but still much greater than one and hence within the range where one can apply the evidence procedure.

Before collecting experimental data and applying the evidence procedure one must reveal one's initial bias $\sigma$.
In this example I assume that there is no prior information available about a specific order of the qubits,
so $\sigma$ must exhibit the symmetry
\begin{equation}
	\langle \vec{S}_i \rangle_\sigma = \vec{s}(\sigma)
	\ 
	\forall i
	\ ,\ 
	\langle C^{ab}_{ij} \rangle_\sigma = {c}^{ab}(\sigma)
	\ 
	\forall i\neq j
	.
\end{equation}
Moreover, there shall be no reason a priori to prefer any particular spatial direction,
an indifference which must be reflected in the initial bias being isotropic:
\begin{equation}
	\vec{s}(\sigma) = 0
	\ ,\ 
	{c}^{ab}(\sigma) = c(\sigma)\cdot \delta^{ab}
	.
\end{equation}
Yet I do assume that -- based on either theoretical considerations or past experience with similar experimental setups -- one has reason to believe that the qubits are correlated, and hence $c(\sigma)\neq 0$.
The initial bias is thus specified by a single parameter characterizing the expected strength of correlations.
This parameter is related to the expected size of the total angular momentum
\begin{equation}
	\vec{J}:=\sum_{i=1}^M \vec{S}_i
\end{equation}
via
\begin{equation}
	\langle \vec{J}^2 \rangle = 3M(M-1)\cdot c + 3M/4
	.
\end{equation}
Since the latter expectation value must lie in the range between zero and $M/2\cdot (M/2+1)$, the strength of correlations is bounded by
\begin{equation}
	-1/4(M-1) \leq c \leq 1/12
	.
\end{equation}
For four qubits, these bounds are $\pm 1/12$.

Given that the initial bias can be characterized completely by the expected magnitude of the total angular momentum, 
it is possible to write $\sigma$ in Maxent form
\begin{equation}
	\sigma = Z(\lambda(\sigma))^{-1} \exp[-\lambda(\sigma) \cdot\vec{J}^2]
\end{equation}
with properly adjusted Lagrange parameter $\lambda(\sigma)$ and partition function
\begin{equation}
	Z(\lambda) := \mbox{tr} \exp[-\lambda \cdot\vec{J}^2]
	.
\end{equation}
For four qubits this partition function can be calculated and all resulting thermodynamic equations solved analytically.
Adding four spins $1/2$ the total angular momentum squared can take the values $j(j+1)$ with $j=0,1,2$.
In the Clebsch-Gordan series these spin-$j$ representations occur with respective multiplicities $m_0=2$, $m_1=3$ and $m_2=1$;
and each such representation has dimension $2j+1$ \cite{cornwell:book2}.
Therefore,
\begin{eqnarray}
	Z(\lambda)
	&=&
	\sum_{j=0}^2 m_j (2j+1) \exp[-\lambda j(j+1)]
	\nonumber \\
	&=& 
	2+ 9\exp(-2\lambda) + 5\exp(-6\lambda)
	;
\end{eqnarray}
whence follows the expectation value
\begin{equation}
	\langle \vec{J}^2 \rangle = - \partial\ln Z/\partial\lambda
	=6\cdot \frac{3\exp(-2\lambda)+5\exp(-6\lambda)}{2+ 9\exp(-2\lambda) + 5\exp(-6\lambda)}
	.
\label{expecvalue}	
\end{equation}
Given the strength of correlations $c$
one can now arrive at the associated $\lambda$ in the following three steps:
relate $c$ to $\langle \vec{J}^2 \rangle$;
then solve the thermodynamic equation (\ref{expecvalue}) for $x:=\exp(-2\lambda)$,
a cubic equation with a unique analytical solution;
and finally obtain $\lambda=-\ln x/2$.
In my example I shall assume a prior bias $c(\sigma)=-0.02$ with associated
$\langle \vec{J}^2 \rangle_\sigma=2.28$, $x(\sigma)\approx 0.7$ and $\lambda(\sigma)\approx 0.18$,
implying $Z(\lambda(\sigma))\approx 10$.

Enter experimental data, measured on a sample of size $N=10,000$.
In order to keep calculations tractable analytically I make the (admittedly artificial) assumption that the $r=48$ different sample means confirm the anticipated symmetries perfectly, and that hence the experimental data, too, can be characterized by the single parameter $c$.
But the measured correlation strength differs from prior expectation, $c\neq c(\sigma)$; 
say, by $25\%$, $c=-0.025$.
Associated with this experimental value is a Maxent state $\mu^\sigma_c$ of the same canonical form as $\sigma$,
yet with different expectation and parameter values $\langle \vec{J}^2 \rangle=2.1$, $x\approx 0.625$ and $\lambda\approx 0.235$,
which implies $Z(\lambda)\approx 8.85$.
This empirical Maxent state deviates from the initial bias by the distance
\begin{equation}
	S(\mu^\sigma_c\|\sigma)=(\lambda(\sigma)-\lambda)\cdot \langle \vec{J}^2 \rangle + \ln Z(\lambda(\sigma)) - \ln Z(\lambda)
	\approx 0.005
	.
\end{equation}
One can now deduce $N_{\rm min}\approx 5,000$ and hence confirm that the two criteria for the applicability of the evidence procedure ($N>N_{\rm min}$ and $r\gg 1$) are indeed satisfied.
From $N\approx 2N_{\rm min}$ results an optimal hyperparameter $\alpha_0$ of approximately the same size as the sample, $\alpha_0\approx N$.
Hence the estimate (\ref{evidenceestimate}) gives approximately equal weight to likelihood and prior;
it features an interpolated Lagrange parameter $\lambda_e\approx 0.208$ which corresponds to
$x_e\approx 0.66$, $\langle \vec{J}^2 \rangle_e\approx 2.19$ and $c_e\approx -0.0226$.

This example illustrates nicely the two distinguishing traits of the evidence procedure.
First, the estimate is not based on experimental data alone;
rather, it takes into account prior information encoded in the initial bias $\sigma$.
In fact, in my example the initial bias and experimental data carried approximately equal weight, so the estimate $c_e$ for the strength of correlations was approximately half way
between its measured value and the prior expectation.
Second, even though in my example the set of sample means was far from informationally complete, the resulting estimate was \textit{just as good} as if it had been based on a set that was informationally complete.
Indeed, since the experimental data were sufficiently close to prior expectations (in particular with regards to the anticipated symmetries) the hyperparameter took a value of similar magnitude as the sample size;
and hence estimates pertaining to degrees of freedom that had not been measured became just as accurate as estimates pertaining to those that had been measured.

\section{\label{summary}Conclusion}

In this paper I transferred the evidence procedure from its historical use in image reconstruction and other  classical estimation tasks 
to a novel use in quantum state tomography.
I verified that this transfer is justified whenever samples are large enough (greater than $N_{\rm min}$) and the set of measured sample means
is sufficiently big ($r\gg 1$), albeit not necessarily informationally complete.
The power of the evidence procedure lies in its optimal use of prior information:
Whenever such prior information is available, the procedure promises to improve both the accuracy and the efficiency of quantum state estimation.
Indeed, I demonstrated in the four-qubit example that the evidence procedure yields a state estimate that is modified with respect to the conventional maximum likelihood or Maxent estimate, by giving some credence still to one's initial bias.
The relative weight attributed to this initial bias is determined by the degree to which the experimental data meet prior expectations overall;
in my example this degree was high because all anticipated symmetries were confirmed perfectly.

In addition to providing a state estimate the evidence procedure also quantifies the associated error bars;
and it does so not only with respect to measured but also with respect to \textit{un}measured degreees of freedom.
Whenever the data are in good agreement with prior expectations overall -- as was the case in my example --  and hence the optimal hyperparameter $\alpha_0$ is of similar magnitude as the sample size $N$,
the unmeasured degrees of freedom are in fact estimated to an accuracy which is just as good as if they had been measured, too.
In situations where the primary goal of an experiment is to confirm or incrementally amend well-founded prior expectations, therefore, 
quantum state estimation becomes very efficient:
Measurement of a limited, not informationally complete set of observables will suffice to estimate the state at a level of confidence extending uniformly to all degrees of freedom,
even the unmeasured ones.
I have not yet considered experimental errors introduced by inaccuracies of the measurement apparatus.
Their treatment should be straightforward conceptually, and will be the subject of future work.

Finally, I speculate whether the evidence procedure may also have some import on foundational issues in statistical mechanics.
The thermodynamical description of macroscopic systems by means of (generalized) canonical distributions always represents a theoretical model which can never be verified in its entirety ---
after all, who would ever verify experimentally, say, the predicted $42$-body correlations in a liquid?
Nevertheless, we use thermodynamical models with a high degree of confidence.
The evidence procedure might provide a justification for this:
As long as the predictions of the thermodynamical model are verified for some limited set of observables,
one has good reason to trust its predictions for all the other observables, too.

\begin{acknowledgments}
I thank Koenraad Audenaert for a stimulating discussion.
\end{acknowledgments}


%

\end{document}